\documentclass[aps,prl,twocolumn,groupedaddress,nofootinbib,10pt]{revtex4-1}
\usepackage{amsmath}
\usepackage{graphicx}
\usepackage{sistyle}
\usepackage{gensymb}
\usepackage{bm}
\usepackage{natbib}
\usepackage{dsfont}
\usepackage{epsfig}
\usepackage{amssymb}

\usepackage{color}

\newcommand{\glass}{SiO$_2$~}

\begin{document}

\title{Optical Rotation of Levitated Spheres in High Vacuum}

\author{Fernando Monteiro}
\email{fernando.monteiro@yale.edu}
\author{Sumita Ghosh} 
\author{Elizabeth C. van Assendelft}
\author{David C. Moore}
\affiliation{Wright Laboratory, Department of Physics, Yale University, New Haven, CT 06520, USA}

\begin{abstract}
A circularly polarized laser beam is used to levitate and control the rotation of microspheres in high vacuum. At low pressure, rotation frequencies as high as 6~MHz are observed for birefringent vaterite spheres, limited by centrifugal stresses.  Due to the extremely low damping in high vacuum, controlled optical rotation of amorphous SiO$_2$ spheres is also observed at rates above several MHz. At $10^{-7}$~mbar, a damping time of $6\times10^4$~s is measured for a $10\ \mu$m diameter SiO$_2$ sphere.  No additional damping mechanisms are observed above gas damping, indicating that even longer damping times may be possible with operation at lower pressure.  The controlled optical rotation of microspheres at MHz frequencies with low damping, including for materials that are not intrinsically birefringent, provides a new tool for performing precision measurements using optically levitated systems.
\end{abstract}

\maketitle

\textit{Introduction.---}Techniques to levitate micron-sized masses in vacuum are under development for use as precision force and torque sensors in a variety of applications~\cite{Barker:2015,Fonseca:2016,Kane:2017,Romero:2017,DUrso:2018,Millen:2018}. Recent work has demonstrated the stable trapping and control of the center-of-mass (COM) motion of optically levitated dielectric spheres in high vacuum~\cite{Ranjit:2015,Ranjit:2016,Jain:2016,Frimmer:2017,Rider:2017,Acceleration_2017}. In addition, control and measurement of angular degrees of freedom~\cite{Bhattacharya:2016} have been demonstrated for optically trapped particles in fluids~\cite{Dunlop:1998,Dunlop:2004,Dunlop:2013}, as well as in air and at moderate vacuum pressure for both torsional~\cite{Hoang:2016,Wang:2004,Baker:2017} and rotational motion~\cite{Dunlop:1995_OAM,Kishan:2011,kishan:2013,Kishan:2016_OAM,Baker:2017,Millen2017,Millen2017_2,Huizhu:18}.

In high vacuum, levitation and rotation of electrically charged, micron-sized graphene flakes has been previously demonstrated~\cite{Kane:2010,Kane:2017}.  However, optical techniques for levitation and rotation of electrically neutral masses have not been studied below $\sim 10^{-3}$~mbar~\cite{kishan:2013}, where feedback cooling of the COM degrees of freedom is necessary to maintain stable trapping. 

Here, a system capable of fully optical levitation and rotation of both birefringent and amorphous dielectric spheres in high vacuum is demonstrated.  Optical levitation of dielectric spheres offers advantages for certain classes of precision sensors since they can be electrically neutralized and their charge controlled at the single $e$ level~\cite{Moore:2014,Frimmer:2017}; they have a highly uniform, spherical geometry; spheres with diameters ranging from $\sim$10~nm to $\sim$20~$\mu$m can be trapped~\cite{Acceleration_2017}; and long working distances ($>$ several cm) between the focusing optics and trap location permit the use of a variety of excitation mechanisms and shielding electrodes~\cite{Ranjit:2015,Rider:2017,Acceleration_2017}.

In this work, the optically driven rotational motion of levitated spheres is observed at frequencies above several MHz, limited by centrifugal stresses.  At $10^{-7}$~mbar, a damping time of $6\times 10^{4}$~s is measured for an amorphous \glass sphere at a rotation frequency above 4~MHz.  At these rotation frequencies and pressures, and in the absence of an externally applied torque, microspheres rotate for $\sim 10^{11}$ cycles in a single damping time due to the extremely low gaseous drag.  No damping mechanisms above drag due to the residual gas are observed, indicating that even lower dissipation may be possible at pressures below $10^{-7}$~mbar.

\begin{figure}[t]
	\centering
	\includegraphics[width=\columnwidth]{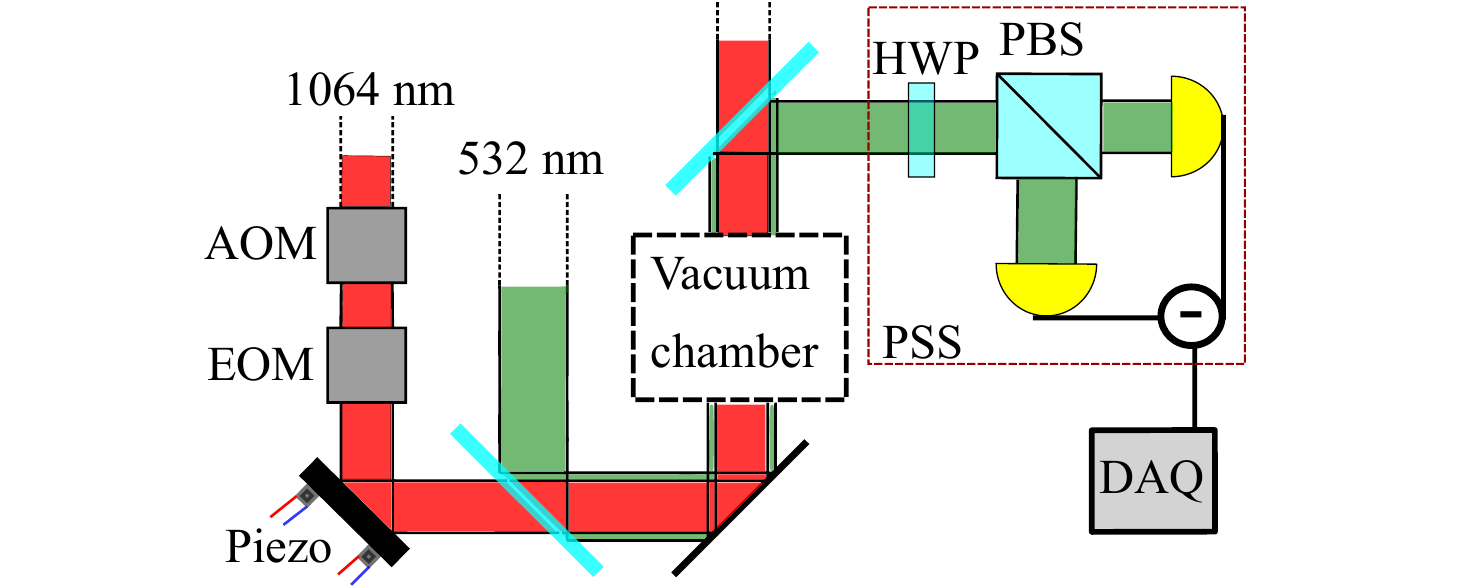}
	\caption{Simplified schematic of the optical setup. The $\lambda=1064$~nm levitating beam passes through an acousto-optic modulator (AOM) and a piezo-controlled mirror, which allow modulation of the power and position of the trapping beam.  An EOM is used to control the beam's polarization. A $\lambda=532$~nm imaging beam with fixed linear polarization is aligned to be coaxial with the trapping beam. The output imaging beam goes to the polarization sensitive sensor (PSS), which sends the signal to the data acquisition system (DAQ). The COM imaging sensors are not shown \cite{Acceleration_2017}.}
	\label{PSS}
\end{figure}

\textit{Experimental setup.---}The experimental setup used here is a modified version of the one reported in a previous work~\cite{Acceleration_2017}, and is depicted in Fig.~\ref{PSS}. An upward propagating trapping beam with wavelength $\lambda = 1064$~nm is used to levitate the microspheres, while two additional beams at $\lambda = 532$~nm with larger waist are used to image the three dimensional motion of the levitated particle. The imaging signals are sent to a field-programmable gate array (FPGA), which controls a feedback loop that damps the motion of the COM degrees of freedom of the sphere in order to maintain stable trapping at low pressure~\cite{Acceleration_2017,Ashkin:1977,Li:2011,Moore:2014}.

The trapping beam also passes through a LiNbO$_3$ electro-optic modulator (EOM) before entering the vacuum chamber. The EOM allows the trapping beam's polarization to be continuously tuned between left and right circular polarization. Accordingly, the optical torque acting on the sphere can be controlled from positive to negative values~\cite{kishan:2013,Dunlop:1998,Millen2017,Millen2017_2}.
The linearly polarized $\lambda = 532$~nm beam aligned along the trapping beam axis also passes through the sphere and is imaged onto a polarization sensitive sensor (PSS) that consists of a polarizing beam splitter (PBS) followed by a balanced photodiode~\cite{kishan:2013,Wang:2004}. A half-wave plate (HWP) is used to balance the power on the photodiodes before trapping the spheres. Since the polarization of the light transmitted through the sphere depends on the angle between the imaging beam polarization and the fast axis of the birefringent sphere, the resulting signal on the PSS is modulated as the sphere rotates around the beam axis~\cite{kishan:2013}.

Two types of microspheres were used in this work: commercially available amorphous SiO$_2$ spheres with diameters of $10.3\pm1.4\ \mu$m~\cite{Acceleration_2017} and vaterite spheres with diameters of $4.9\pm0.47\ \mu$m. Vaterite is a polymorph of CaCO$_3$ for which polycrystalline birefringent spheres can be grown~\cite{vaterite_structure}. The vaterite microspheres used here were fabricated by agitating a mixture of 25 mL 0.1 M CaCl$_2$, 1 mL 0.1 M MgSO$_4$, and 1.5 mL 0.1 M K$_2$CO$_3$~\cite{vaterite_parkin_09} until spheres were suspended in the residual salt solution. The spheres were separated from the residual solution by centrifugation and stored in ethanol for several weeks without degradation~\cite{wateraffectsthevaterite}.

\begin{figure}[t]
	\centering
	\includegraphics[width=\columnwidth]{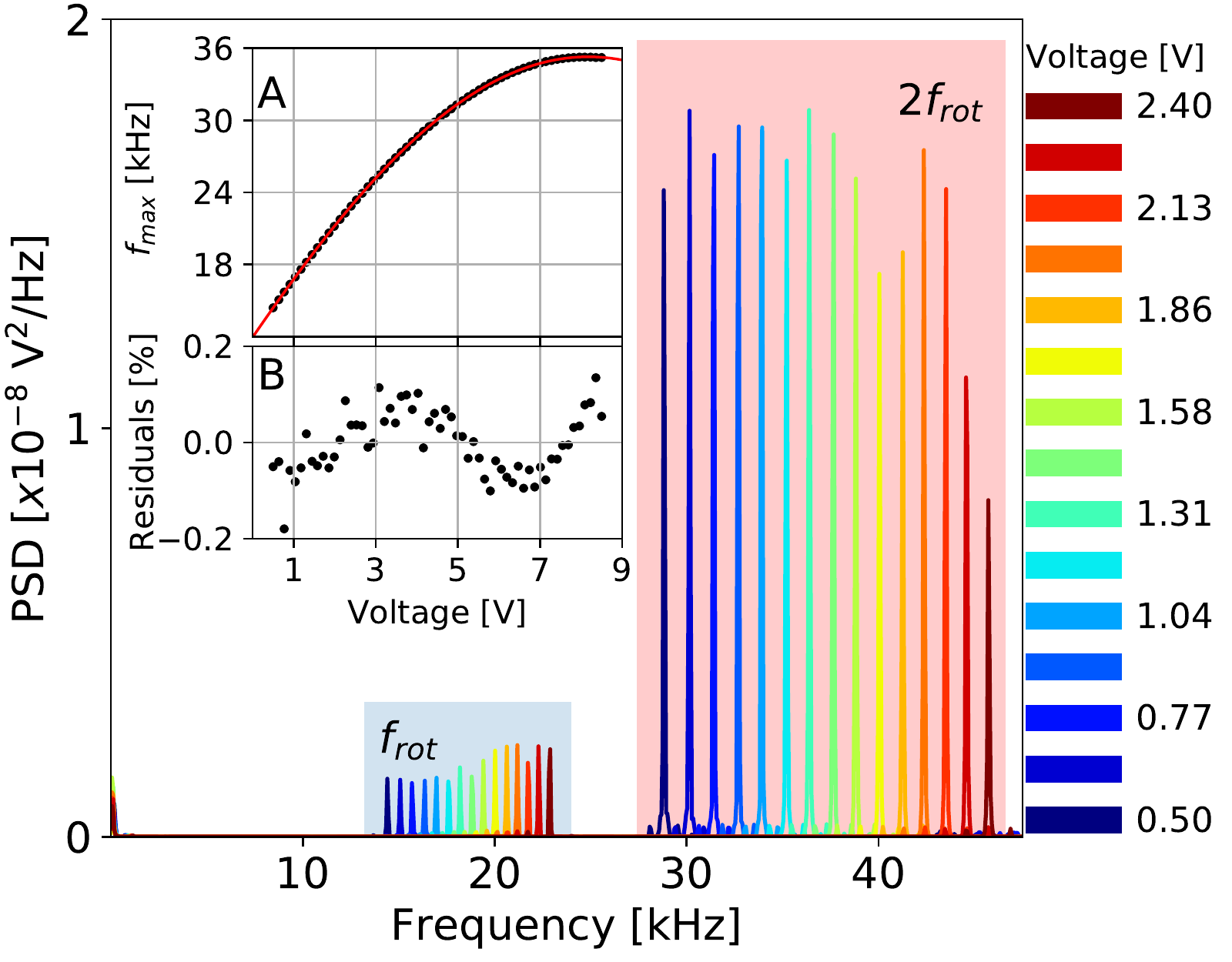}
	\caption{Power spectral density (PSD) of the signal measured by the PSS for a 5~$\mu$m diameter vaterite sphere. Data is acquired at a constant pressure of $\sim4\times10^{-2}$ mbar with varying trapping beam polarization. Peaks at both $f_{rot}$ and $2f_{rot}$ are observed as described in the text. Inset A): Measured terminal rotational frequency, $f_{max}$, as a function of the EOM voltage (black dots) and best-fit model (red line). B): Relative fit residuals, $(data-fit)/data$.}
	\label{fig:vaterite_004mbar}
\end{figure}

\textit{Results.---}The rotation frequency, $f_{rot}$, for both vaterite and \glass spheres was measured at various pressures and EOM settings. Figure~\ref{fig:vaterite_004mbar} shows the power spectrum of the PSS signal for a trapped vaterite sphere at pressure $p \sim 4\times10^{-2}$~mbar as the control voltage applied to the EOM was varied. For each EOM setting, two peaks are observed in the spectrum. The largest peak corresponds to the polarization modulation caused by rotation of the birefringent sphere as described above, and is positioned at a frequency of $2f_{rot}$~\cite{kishan:2013}. A smaller peak is observed at $f_{rot}$, likely resulting from modulation of the power reaching the sensor due to residual asphericity of the microsphere. Higher harmonics resulting from this asphericity are also observed, although typically with smaller amplitudes than the peak at $f_{rot}$.

By identifying the location of the peaks at $f_{rot}$ and $2f_{rot}$ for each EOM voltage, the dependence of the terminal rotational frequency, $f_{max}$, for a given trapping beam polarization can be measured.  The terminal frequency is reached at a given pressure when the torque due to gaseous drag balances the optical torque.  Fig.~\ref{fig:vaterite_004mbar}(A) shows the measured $f_{max}$ as a function of EOM voltage. The optical torque, $N_{opt}$, on a spinning birefringent sphere is given by $N_{opt} = a\sin[b(V-V_0)]$~\cite{Dunlop:1998,Car:2001}, where the parameter $a$ is related to the sphere-dependent birefringence and the optical power, $b$ is related to the EOM's Pockels coefficient, and $V_0$ is an offset voltage at which the net optical torque is zero. Since the torque due to gaseous drag is $N_{drag} \propto f_{rot}$~\cite{FREMEREY_1982,Fremerey_1985}, $f_{max}$ is proportional to $N_{opt}$. For the data shown in Fig.~\ref{fig:vaterite_004mbar}, a delay of several seconds was incorporated between setting the EOM's voltage and measuring $f_{max}$. This delay is much longer than the expected damping time at this pressure, calculated to be $\sim 0.2$~s~\cite{FREMEREY_1982,Fremerey_1985}, and consistent with the measured damping times presented below. This model for the terminal rotation speed is in good agreement with the measured $f_{max}$, and the relative residuals obtained from the fit and shown in Fig.~\ref{fig:vaterite_004mbar}(B) are $\lesssim 0.1\%$. The best-fit offset is $V_0 = -2.3$~V, consistent with the specified extinction ratio of $10$~dB for this EOM when operated as an amplitude modulator.

\begin{figure*}[t]
	\centering
	\includegraphics[width=\textwidth]{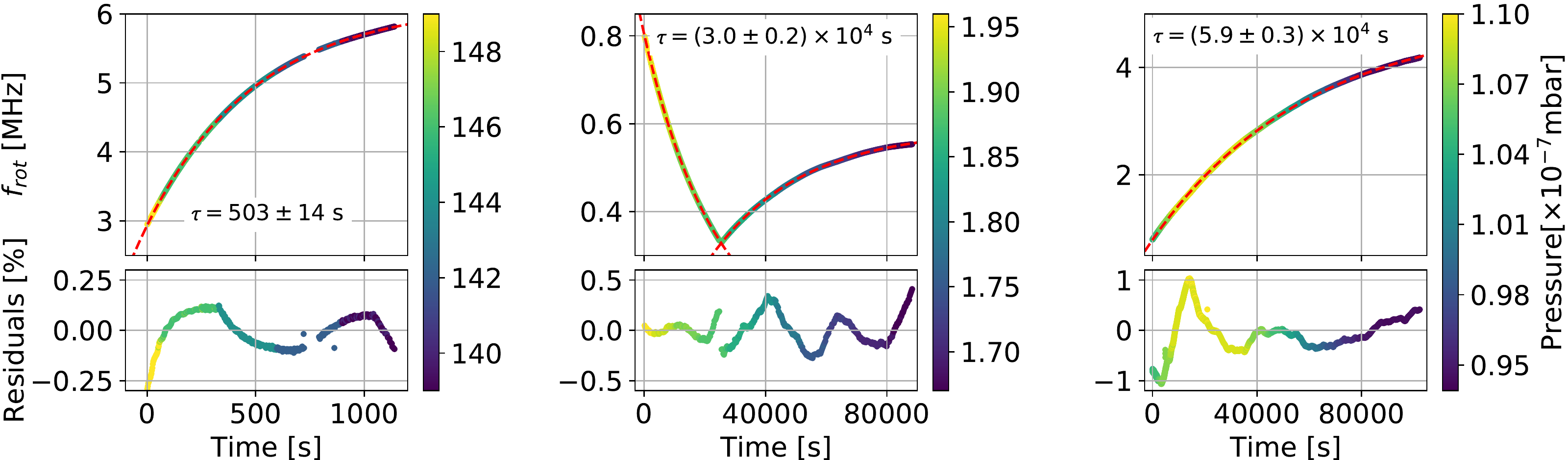}
	\caption{Damping time measurements of a vaterite sphere (left) and two amorphous SiO$_2$ spheres (center and right) along with their relative residuals when fit to an exponential curve (red, dashed). Both the data points and residuals are colored by the pressure recorded on the cold cathode gauge in the vacuum chamber, indicated by the color scale to the right of each plot. The turning point in the center plot is where the direction of the torque applied by the EOM was reversed.}
    \label{fig:3plots}
\end{figure*}

The terminal speed was also measured at pressures from $\sim 10^{-3}$ to $\sim 10^{-5}$~mbar by introducing a small N$_2$ flow through a leak valve while pumping with a turbo-molecular pump set at a low rotation speed. At lower pressures, vaterite spheres quickly spin up to rotation speeds $f_{rot} > 5$~MHz, where the centrifugal stress approaches the expected yield strength of the material, and spheres are consistently lost from the trap~\cite{kishan:2013}. As shown in Fig.~\ref{fig:3plots}~(left), a constant applied optical torque at constant pressure causes a vaterite sphere to accelerate to a rotational frequency near 6~MHz before it is lost. The rotational frequency versus time, $t$, is fit by an exponential curve, $f_{rot}(t) = f_{max}(1 - e^{-(t-t_0)/\tau})$, where $t_0$ is a time offset and $\tau$ is the $1/e$ damping time. The damping time determined from the fit is $\tau = 503\pm14$~s at $p \sim 1.4\times10^{-5}$~mbar, where the error on $\tau$ is dominated by pressure fluctuations. The applied optical torque can be determined from $N_{opt} = I\omega_{max}/{\tau}$ where $I$ is the moment of inertia and the terminal angular frequency, $\omega_{max} = 2\pi f_{max}$, is determined from the fit. Following this procedure, the optical torque is found to be $N_{opt} = 28\pm13$~fN$\,\mu$m, with the error primarily due to the uncertainty on the sphere's radius and moment of inertia. The optical angular acceleration $\alpha_{opt} = \omega_{max}/{\tau}$ is independent of the sphere's radius and is found to be $\alpha_{opt} = (76.8\pm1.5)\times10^{3}$~rad~s$^{-2}$. This optical torque and angular acceleration result from a nearly linearly polarized beam, for which only a small amount of residual ellipticity is sufficient to quickly increase $f_{rot}$, resulting in the loss of the microsphere. As described below, an order of magnitude larger torque can be applied for vaterite spheres when circularly polarized light is used. Future work will allow the optical torque to be controlled by the feedback system, enabling controlled rotation of vaterite spheres at lower pressure.

Although not expected to exhibit substantial birefringence, rotation of amorphous SiO$_2$ spheres was also observed at low pressure by the presence of large peaks in the power spectrum of the signal recorded by the PSS, which increased in frequency as the pressure was reduced. As for the vaterite spheres, the PSS spectrum contains peaks located at $f_{rot}$, $2f_{rot}$, and higher harmonics. For the \glass spheres, the relative amplitude of the peak at $2f_{rot}$ varies from being roughly equal to that at $f_{rot}$ to being up to $\sim$10$\times$ larger, depending on the sphere.  This variation could indicate that the PSS's sensitivity to rotation is primarily due to sphere-dependent asphericity. The transfer of angular momentum from the trapping laser to the sphere could arise from either trace amounts of birefringence, small amounts of asphericity, or absorption of the trapping light.

As shown in Fig.~\ref{fig:3plots}~(center), it is possible to control the direction of the angular acceleration by changing the voltage applied to the EOM. Similar to the vaterite spheres, a microsphere dependent offset to the voltage applied to the EOM was necessary to minimize the angular acceleration of the \glass spheres. However, unlike the vaterite spheres, this offset can be sufficiently large that no EOM setting is able to reverse the acceleration direction. Such an offset torque could arise, e.g., from absorption of light that has small components of orbital angular momentum~\cite{Dunlop:1995_OAM}, possibly due to misalignment of the levitation and imaging beams~\cite{astig:1993,astig:2004,astig:2005,Huizhu:18}.

In order to isolate the offset torque from the EOM-dependent torque, the change in $f_{max}$ was measured after changing the trapping beam polarization from approximately linear to circular. The maximum torque that can be provided by the EOM was found to be $N_{opt} = 2.6\pm1.7$~fN$\,\mu$m for the sphere shown in Fig.~\ref{fig:3plots}~(right), corresponding to an angular acceleration of $\alpha_{opt} = (234\pm5)$~rad~s$^{-2}$. This acceleration could arise from an absorption of $\sim 10^{-5}$ of the trapping laser power, consistent with the absorption expected for water impurities within the microspheres~\cite{water_absorption_hale1973optical,Acceleration_2017}.

The measured damping time for the SiO$_2$ sphere in Fig.~\ref{fig:3plots}~(right) is $\tau = (59.4\pm 3.3)\times10^{3}$~s at $\sim 10^{-7}$~mbar. To determine the damping time, the measured rotation frequency versus time is fit to the same function used previously for the vaterite sphere. This fitting function does not account for possible drifts in pressure during the measurement.  However, modification of the fitting function~\cite{FREMEREY_1982} to incorporate the measured pressure at a cold cathode gauge located $\sim$10~cm from the trap location was found to consistently lead to larger residuals than for the simple exponential fit.  This increase in residuals likely indicates that the pressure gauge does not accurately reflect the pressure fluctuations surrounding the sphere, and that any residual fluctuations are small relative to the precision of the gauge. The measured rotation is thus expected to provide a more accurate constraint on the pressure in the microscopic region around the sphere. Nevertheless, the measured pressure fluctuations using the cold cathode gauge are conservatively included when determining the corresponding errors for the measurements of the damping times in this work.

\begin{figure}[t]
	\centering
	\includegraphics[width=\columnwidth]{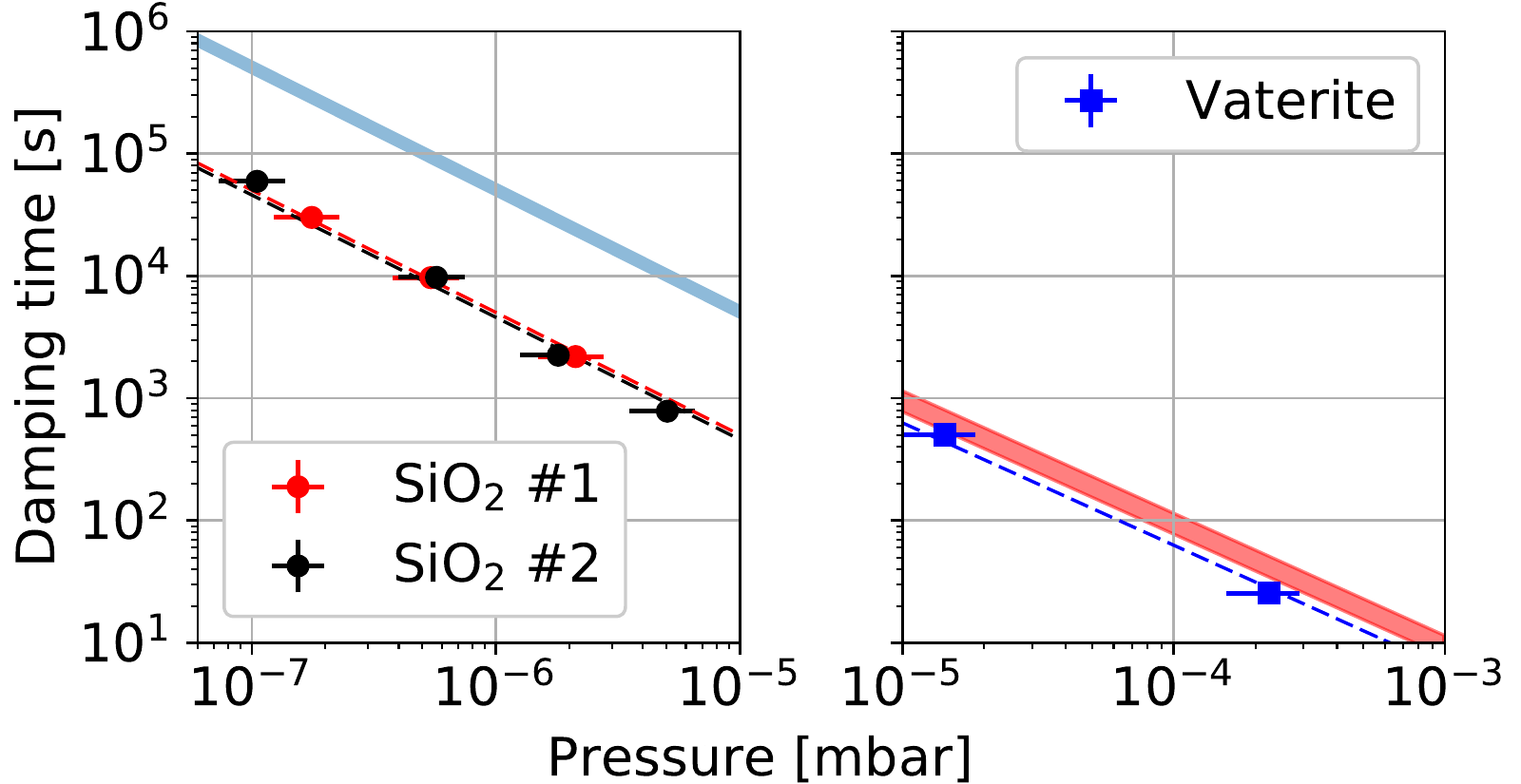}
	\caption{Damping time vs pressure for the two \glass spheres (left) and the vaterite sphere (right) measured here. The blue and red bands above the data points indicate the expected damping times for $\sigma=1$, while the red, black, and blue dashed lines are the best fit to a $1/p$ dependence. The horizontal error bars indicate the systematic error of the cold cathode pressure gauge.}
    \label{fig:damping_time}
\end{figure}

Following the procedure above, the damping time was measured at several pressures for one vaterite and two SiO$_2$ spheres, as shown in Fig.~\ref{fig:damping_time}. 
The dashed lines indicate the best fit $1/p$ dependence, which is expected if the residual gas provides the primary damping mechanism~\cite{FREMEREY_1982}. No decrease in the damping time relative to the $1/p$ dependence is observed down to pressures $\lesssim 10^{-7}$~mbar, indicating that any additional damping mechanisms are sub-dominant.  This is in contrast to the COM degrees of freedom for levitated microspheres, where technical sources of noise typically dominate dissipation due to residual gas at comparable pressures~\cite{Ranjit:2015,Ranjit:2016,Acceleration_2017}.

The expected damping times arising from gaseous drag can be calculated for a rotating sphere in the molecular flow regime following~\cite{FREMEREY_1982} as $\tau = \pi\rho \overline{c}r/(10\sigma p)$, for pressure $p$, mean molecular velocity $\overline{c}$, sphere radius $r$, and sphere density $\rho$. Here, $\sigma$ is the empirical accommodation coefficient, which parameterizes details of the surface roughness and composition of the spheres~\cite{FREMEREY_1982}. The predicted damping time (assuming $\sigma = 1$) is indicated by the filled bands in Fig.~\ref{fig:damping_time}. The blue band on the left plot of Fig.~\ref{fig:damping_time} assumes the molecular velocity for H$_2$, since a residual gas analyzer was used to verify that H$_2$ was the dominant gas species after the chamber was initially evacuated to the base pressure $\lesssim 10^{-7}$~mbar, and when subsequent measurements at higher pressure were performed by reducing the turbo pump rotation speed. The red band of the right plot assumes a gas composition that is predominantly N$_2$, since a small flow of this gas was introduced into the chamber to control the pressure between $\sim 10^{-3}$ to $\sim 10^{-5}$~mbar.

The measured damping times for the \glass and vaterite spheres correspond to $\sigma^{-1} \approx 0.1$\textendash0.5.  These damping times, which are significantly shorter than predicted by the simple model above, could be attributed to the detailed interaction of the gas molecules with the surface of the microspheres.  While macroscopic spinning rotor pressure gauges typically have $\sigma \approx 1$, surface roughness or non-metallic coatings are known to lead to a non-unity accommodation coefficient~\cite{FREMEREY_1982}. In particular, surface effects for microscopic objects should have a substantially larger effect than for macroscopic objects due to the high surface-to-mass ratio.  It is also possible that the surface temperature of the sphere is elevated from the gas temperature~\cite{Acceleration_2017}, but this would not be expected to significantly reduce the damping time~\cite{Stickler:2018}. Alternatively, the shorter damping times could arise if there were a substantial difference between the pressure in the vicinity of the sphere and that measured by the cold cathode gauge. However, if such pressure differential were present, it would be required to be stable within the sub-percent residuals of the exponential fit over the $10^5$~s long measurements.

The measured dependence of the damping time with pressure can be used to determine the maximum torque applied on the vaterite sphere at higher pressure, where circularly polarized light can be used without loss of the sphere. The extrapolated damping time for a vaterite sphere at $4\times10^{-2}$~mbar is found to be $\leqslant 0.2$~s. Using the $f_{max}$ obtained for this sphere in Fig.~\ref{fig:vaterite_004mbar}, a maximum torque of $N_{opt} \geqslant 300$~fN$\,\mu$m and angular acceleration of $\alpha_{opt} \geqslant 10^{6}$~rad~s$^{-2}$ are obtained. The maximum angular acceleration is at least 3 orders of magnitude larger for the vaterite spheres than for the SiO$_2$ spheres. Accordingly, although even small amounts of torque are sufficient to rotate spheres at MHz frequency in high vacuum, for applications that require the maximum possible applied torque, vaterite or other birefringent materials can be used. 

\textit{Conclusion.---}This work has demonstrated control over the rotation of optically levitated vaterite and amorphous \glass spheres at pressures as low as $10^{-7}$~mbar. At low pressures, the rotation frequency can exceed several MHz for both types of spheres, limited by centrifugal stresses. The angular acceleration that can be applied to the vaterite spheres is 3 orders of magnitude larger than for the \glass spheres, allowing MHz frequencies to be reached at pressures as high as $10^{-3}$~mbar. As the pressure decreases, the rotation frequency for vaterite spheres quickly exceeds $\sim5$ MHz, leading to their loss from the trap~\cite{kishan:2013}. Future work will incorporate a feedback system capable of controlled rotation of birefringent spheres at lower pressures, including the possibility to lock the rotation to a given frequency~\cite{Millen2017, Millen2017_2}. 

The rotational damping time is found to be inversely proportional to the pressure, and is measured to be $\sim 6\times 10^{4}$~s at $10^{-7}$~mbar for a $10\ \mu$m \glass sphere.  The damping time versus pressure is well-described by a $1/p$ dependence down to the base pressure of the system, indicating that larger damping times may be possible at lower pressure.  

The rotational degrees of freedom of optically levitated microspheres may offer an extremely low dissipation system for precision force or torque sensors~\cite{Hoang:2016,Millen2017_2}. Controlled rotation of microspheres in high vacuum may allow the reduction of backgrounds that couple to electric dipole moments in the spheres~\cite{Moore:2014,Rider:2016}, precision measurements of torques on gyroscopically stabilized rotating spheres~\cite{kishan:2013,Kane:2017}, micron-scale pressure measurements in high vacuum~\cite{FREMEREY_1982}, or possibly even tests of predictions of fundamental mechanisms of dissipation for rotating systems~\cite{Abajo:2010,Abajo:2012,Abajo:2017}. Finally, the demonstration that even non-birefringent materials can be rotated at MHz frequencies in high vacuum can enable the broad application of these techniques to a variety of optically levitated systems.

\textit{Acknowledgements.---}We would like to thank C.~Blakemore, G.~Gratta, A.~Kawasaki, and A.~Rider (Stanford University) for useful discussions related to this paper. This work is supported, in part, by the Heising-Simons Foundation, the National Science Foundation under Grant No. 1653232, and Yale University.

\bibliographystyle{apsrev4-1}
\bibliography{references}{}

\end{document}